\documentclass{epl}

\usepackage{amsmath}
\usepackage{amssymb}
\usepackage{graphicx}

\title{Simulations of collision times in gravity driven granular flow}
\author{John J. Drozd and Colin Denniston}

\institute{                    
Department of Applied Mathematics, The University of Western 
Ontario, London, Ontario N6A 5B8, Canada\\
}

\begin{document}

\maketitle

\begin{abstract}
We use simulations to investigate collision time distributions as one approaches the static limit of steady-state flow of dry granular matter. The collision times fall in a power-law distribution with an exponent dictated by whether the grains are ordered or disordered.  Remarkably, the exponents have almost no dependence on dimension.  We are also able to resolve a disagreement between simulation and experiments on the exponent of the collision time power-law distribution.
\end{abstract}

Dense granular matter does not easily fit into our standard classification of matter.  Interest in the physics community was stimulated by the observation that forces between particles in the system are exponentially distributed suggesting that any load on the system is carried by a small number of force chains \cite{Jaeger96,Mueth}. More recent studies have suggested that spatial ordering is a key factor in the force response \cite{Behringer01,Donev}, something that may not be clearly distinguishable from the exponential tail of the force distribution.  Further work, both experimental and theoretical, has suggested that perhaps it is actually the low end of the force distribution that should be examined when deciding whether a system is jammed \cite{OHernandNagel,Ferguson,Silbert}.

As the distribution of forces in a static granular system is history dependent \cite{Tsai04}, it makes sense to explore the static limit of dynamic models in order to shed light on the origin of jamming.  In this letter, we perform simulations of gravity driven dense granular flow in two and three dimensions using an event-driven model involving only binary collisions.  We reproduce the known power law relationship between the mean flow velocity and velocity fluctuations \cite{Menon} and explain the discrepancy between the experimentally observed power law for the distribution of collision times \cite{Longhi} and that found in previous simulations \cite{Denniston}.  Further, we find that the distribution of collision times is capable of differentiating between ordered and disordered glassy systems.  Moreover, the exponents appear to be superuniversal in the sense that they are independent of dimension. In addition, we find that the force distribution turns {\it up} at low impulses in the jammed state.

\begin{figure}
\includegraphics[width=12.5cm]{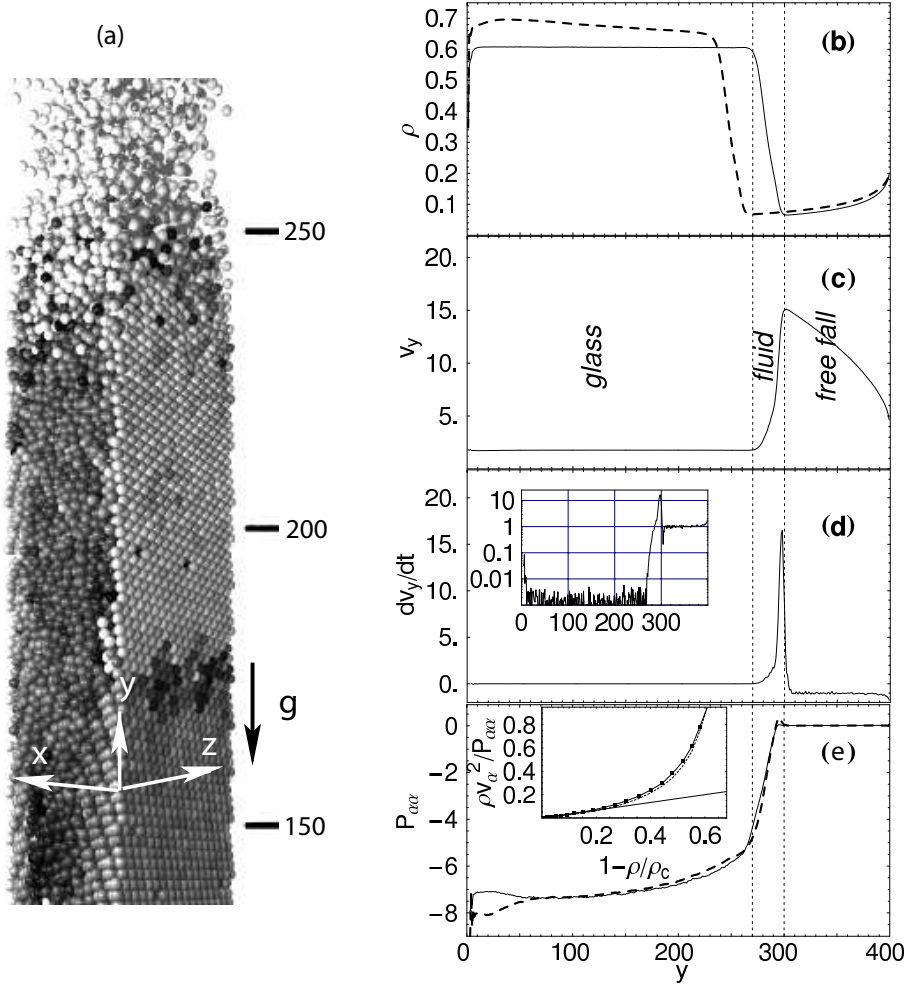}
\caption{(a) Section of a simulation involving $43\,200$ grains 
with $1\,\%$ polydispersity in a $32a \times 32a \times 400a$ system.
Average (b) density for 1\% (dashed line) and 15\% polydispersity (solid line) along the height of a 3D chute.  For 15\% polydispersity,
(c) the $y-$velocity and (d) average acceleration, in units of $g$ and as measured by the material derivative $dv_y/dt=\partial_t v_y+v_\alpha \partial_\alpha v_y$.    
The inset in (d) shows the absolute value of the same data on
a log scale, showing the $ 1 g$ acceleration in the free fall
region.  (e) shows the pressure tensor components $P_{xx}$ as the solid line and $P_{yy}$ as the dashed line.  The inset in (d) shows that the pressure approaches Eq. (\ref{pressure}) as jamming is approached from the fluid phase\cite{Donev}.}
\label{fig1}
\end{figure}

We simulate hard spheres as they fall down a rectangular chute under the influence of gravity, as shown in Fig.~\ref{fig1}(a). At the bottom of the chute a sieve is modeled by having the particles reflecting from the bottom with a probability $p$ (typically $p=10 \%$).  Particles transmited through the bottom reappear at the top of the chute once again to fall down through it.  Particles reflect off the walls of the chute with a partial loss, typically 10 $\%$, in their vertical velocity.  This is a simple effective way to model rough walls.  The grain polydispersity is varied from $0$ to $15\%$ in different simulations \cite{polydispersity}.

Velocities after collision $\dot{\bf r}_1'$ and $\dot{\bf r}_2'$ in terms of
velocities before collision, $\dot{\bf r}_1$ and $\dot{\bf r}_2$, are
\begin{equation}
\left( \begin{array}{c} \dot{\bf r}_1' \\ \dot{\bf r}_2'
\end{array} \right)= \left( \begin{array}{c} \dot{\bf r}_1 \\
\dot{\bf r}_2 \end{array} \right)+ \frac{(1+\mu)}{(m_1 + m_2)} \left(
\begin{array}{cr} -m_2 & m_2 \\ m_1 & -m_1 \end{array} \right) \left(
\begin{array}{cc} \dot{\bf r}_1 \cdot {\bf q} \\ \dot{\bf r}_2
\cdot {\bf q} \end{array} \right){\bf q},
\label{collide}
\end{equation}
where ${\bf q}=({\bf r}_2-{\bf r}_1)/|{\bf r}_2-{\bf r}_1|$ \cite{JenkinsSavage,SavageJeffrey}.  $\mu$ is a velocity-dependent restitution coefficient
described by the phenomenological relation\cite{Swinney,Luding96},
\begin{equation}
\mu\left(v_n\right)=\left\{ 
\begin{array}{ll}
  1-\left(1-\mu_0\right)\left(v_n/v_0\right)^{0.7} &,\, v_n<v_0 \\ 
  \mu_0 &,\, v_n>v_0.
\end{array} 
\right. 
\label{eqn2}
\end{equation}
Here $v_n$ is the component of relative velocity along the line
joining the grain centers, $\mu_0$ is the asymptotic coefficient at large velocities (typically 0.9 here), and $v_0=\sqrt{2ga}$, with $g$ being the acceleration due to gravity and $a$ the radius of the particle \cite{velocity}.

We examine three simulation geometries: ``disks'', a purely two-dimensional simulation with disks as used in \cite{Denniston}; ``spheres in 2D'' which are spherical particles in a three-dimensional (3D) simulation kept in a 2D plane by having reflecting front and back walls $2.002 a$ apart and with reflecting left and right walls; and ``spheres in 3D'' which are spherical particles in a chute with periodic front and back walls $16-32 a$ apart and with reflecting left and right walls.  In all three types of simulations a sieve is located at the bottom as described above.  For simulations it is most convenient to use units where $g=1$, $a=1$, and the grain mass $M=1$.  A typical 
simulation run is $10^4-10^5$ time units, equivalent to $2-20$ minutes of
real time in a system made up of balls $3$ mm in diameter.

Steady-state density and velocity profiles for a system similar to that 
depicted in Fig.~\ref{fig1}(a) are shown in Fig.~\ref{fig1}(b) and (c).   
These profiles
 are averaged over the $xz-$cross-section of the $32a\times 32a$ chute.  Three 
different regimes are evident.  
Grains initially placed at the top are little affected by collisions and accelerate 
uniformly with ${\dot v_y}=-g$, as shown in Fig.~\ref{fig1}(d).  This free-fall
region ends abruptly just above the top of the dense column of grains seen in 
Fig.~\ref{fig1}(a).  There is then a short fluid 
region extending about $25a$ between the free-fall and glassy region.  The 
grains behave like a simple fluid in this region, with a parabolic profile of 
the $v_y$ velocity typical of Poiseuille flow ($\square$'s in Fig. 2(b)).  
For nearly monodisperse systems, the density increases rapidly in this region until it reaches the random close-packed volume fraction of $\sim 0.64$.  Except for a 
small region at the bottom, the remainder of the system has a 
density greater than $0.64$ indicating partial crystallization.  For systems of
particles with $15$\% polydispersity, the density remains constant at $0.6$ and is in a glass-like state.  As a result we will now concentrate on $15$\% polydisperse systems as they are more uniform.  

\begin{figure}
\includegraphics[width=12cm]{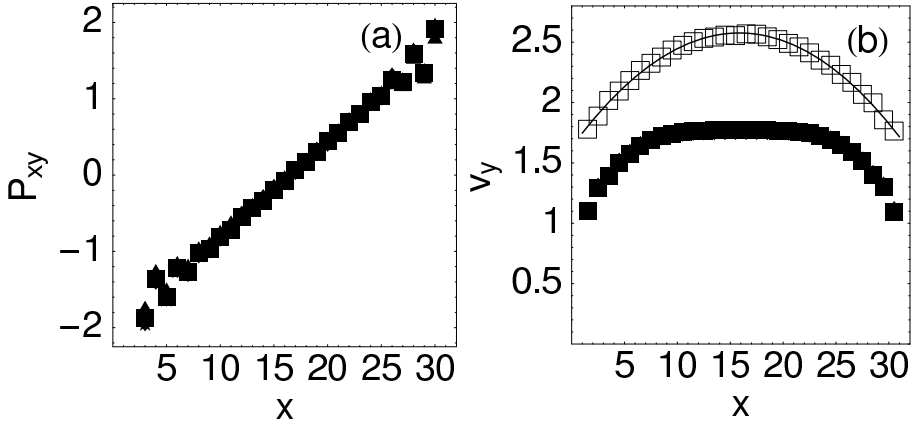}
\caption{Shear stress and velocity profiles for the 15\% polydisperse 3D simulation.  Data in the glassy region for $h=$50, 100, 150, 200 are plotted 
separately but fall on top of each other ($\blacksquare$'s). The $\square$'s in
(b) show the velocity profile in the fluid region at $h=280$ which gives a parabolic profile (solid line).}
\label{fig2}
\end{figure}

Fig.~\ref{fig2}(b) shows the $v_y$ plug profile ($\blacksquare$'s) along with the corresponding shear stress profile in the glassy region of such a system.  Clearly, the system supports a
finite shear stress in the central region thus justifying calling this a glass.
In steady state, the weight of the system, $\rho g$ can be supported by 
either a pressure gradient $\partial_y P_{yy}$ or by a gradient in the shear stress $\partial_x P_{xy}$ (pressures and stresses are measured as in \cite{Denniston}).  A comparison of 
$\partial_y P_{yy}$ and $\partial_x P_{xy}$ (cf. Fig.s~\ref{fig1}(e) and ~\ref{fig2}) shows that the shear stress, and hence the walls, carry most ($\sim 98\%$) of the weight. 

As jamming is approached, Donev et al. \cite{Donev} showed that the pressure in a classical (conservative) hard sphere glass approaches
\begin{equation}
P = D(\rho k_B T)(1-\rho/\rho_c)^{-1},
\label{pressure}
\end{equation}
where $D$ is the dimension, $k_B$ is Boltzmann's constant, and $\rho_c$ is the close-packed density.  As can be seen in the inset in Fig.~\ref{fig1}(e), the diagonal components of the pressure tensor do approach this value as $\rho$ approaches $\rho_c$.  However, there are small systematic deviations in the glassy phase, probably due to the dissipation and the fact that the velocities are not Maxwell distributed \cite{Moka05} as assumed in the derivation of Eq.~\ref{pressure}.  

\begin{figure}
\includegraphics[width=10.0cm]{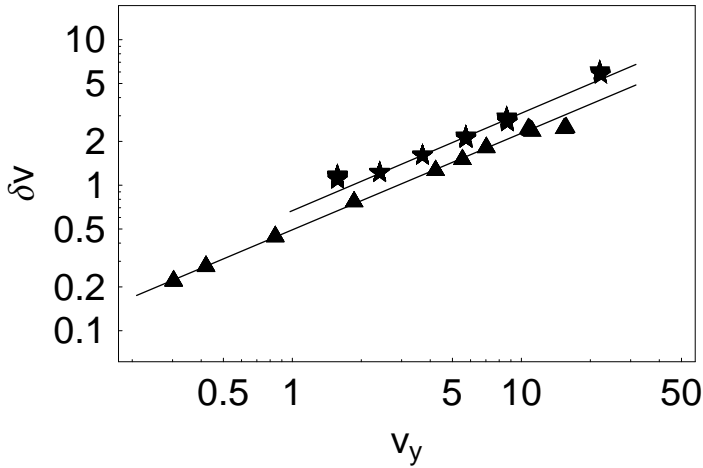}
\caption{Relationship between fluctuating and flow velocity in the glassy region.  Data was averaged in directions normal to $\vec{\mathbf g}$ for the 32x32 ($\star$, $\delta v = \left(\delta v_x^2+\delta v_y^2+\delta v_z^2\right)^{1/2}$) and 16x16 ($\blacktriangle$, $\delta v = \delta v_y$) $15$\% polydisperse systems. The fitted lines have slope of 2/3, in agreement with the experiments of \cite{Menon}.  $v_y$ is varied by varying the probability of reflection $p$ from $0.01$ to $0.995$.}
\label{fig3}
\end{figure}

It is reasonable to doubt the capability of a simple binary collision model to describe the dense glassy phase \cite{Gollub}.  However, it is capable of reproducing a number of experimental results related to this dense phase.  For example, experiments have shown \cite{Menon} that the fluctuating velocity $\delta {\bf v}$ is related to 
the flow velocity ${\bf v}$ by a power-law relationship $\delta {\bf v} \propto {\bf v}^{2/3}$. We observe the same relationship (Fig.~\ref{fig3}).  Velocity fluctuations are not, however, isotropic.  In a equilibrium fluid, fluctuations in the velocity are governed by equipartition and $\delta v_y^2=\langle v_y^2\rangle-\langle v_y\rangle^2$ is the same as $\delta v_x^2$ and $\delta v_z^2$.  This is very nearly the case in what we label the fluid region, but 
is not at all the case in the unequilibrated free fall region or in the glassy
region.  Further measurements, such as that of the kurtosis of
the velocity distribution, showed that the distribution of velocities is not Boltzmann-distributed. 
 
\begin{figure}
\includegraphics[width=14.25cm]{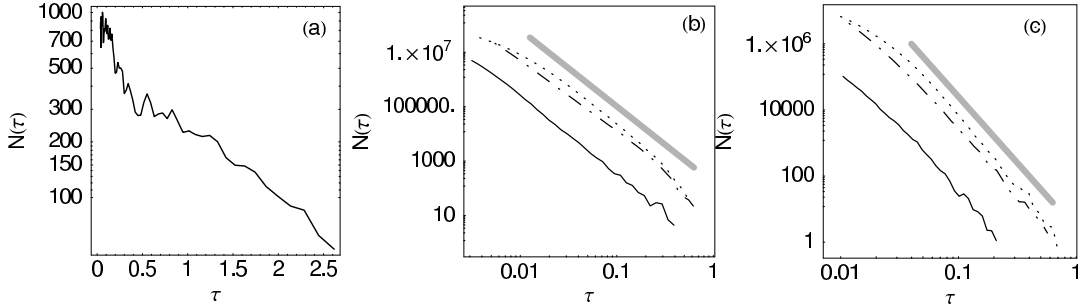}
\caption{(a) Unnormalized collision time distribution in the fluid region of a $15\%$ polydisperse ``3D spheres'' simulation showing an exponential distribution. (b) Unnormalized collision time distributions in the center of the glassy region for  {\it $15\%$ polydisperse} ``disks'' (dash-dotted line), ``spheres in 2D'' (dotted line), and ``3D spheres'' (solid line) showing a power-law with exponent $2.85$ as indicated by the thick gray line.  (b) Similar to (a) but for {\it monodisperse} particles that have crystallized.  The power-law in this case has exponent $4$ as indicated by the thick gray line.   Note, plots in (b) and (c) are log-log plots whereas (a) is a {\it semi}-log plot.
All distributions are averaged over time in a region slightly larger than a grain size in the $x$ and $y$ directions and over the entire $z$ direction.  The precise choice of the location, other than it being clearly in the appropriate region, does not change the result.}
\label{fig4}
\end{figure}

Considerable attention has also been focused on relating the force/impulse and collision time distributions to jamming. A collision time is defined as the time between two consecutive collisions for a particle.
In a simple fluid the collision times are distributed exponentially and the 
mean collision time (or ``relaxation time'') is proportional to the mean
free path, the typical distance traveled between collisions.  This is indeed
what we observe in what we call the ``fluid'' region of the simulations, as shown in Fig.~\ref{fig4}(a).  The distribution
of collision times can be used to test the validity of many of the assumptions
that go into standard kinetic theories used to describe granular gases and 
show us how they break down as we move into the dense glassy regions.  As we
will show, the distribution can also distinguish between a disordered glass and
a crystalline packing of grains.

The distributions of the collision times $N(\tau)$ for binary collisions are shown in Fig. \ref{fig4}.  For the glassy region of the $15\%$ polydisperse systems (Fig.~\ref{fig4}(b)), $N(\tau)$ follows a power law,
\begin{equation}
N(\tau) \sim \tau^{-\alpha}\label{powerlaw}
\end{equation}
where $\alpha$ is $2.81 \pm 0.06$ for distributions of grain-grain collisions
in all the geometries, {\it regardless of dimension}.   The power-law behavior of $N(\tau)$  is independent of the specific choice made for the coefficient of restitution (as long as it is less than 1), and whether or not we use a constant or velocity-dependent coefficient of restitution as in Eq.~\ref{eqn2}.  We also examined the distribution of the distance particles travel between consecutive collisions.  It has the same power-law behavior as the collision time distribution and we will therefore concentrate on the collision time distribution here.  

Experiments often refer to systems with less than $5\%$ polydispersity as monodisperse.  However, truly monodisperse systems eventually crystallize and give a power law with $\alpha=4$ as shown in Fig.~\ref{fig4}(c) (3D monodisperse systems did not crystallize over the entire system so the 3D data in Fig.\ref{fig4}(c) is taken from a crystallized region).  However, we find significant differences between even $1\%$ polydisperse and truly monodisperse systems and systems with $7.5 \%$ polydispersity yielded the same power laws as the $15\%$ system.  It appears that the polydispersity is a critical factor in these calculations mainly due to its affect on breaking up crystalization.    

The significance of the power-law distribution of collision times lies in
the fact that it implies that collisions are not statistically
independent events.  If they were independent, it is 
straight-forward to show that the distribution of collision times would
be exponential, or at least it would have an exponential tail \cite{Reif}.
This does not necessarily suggest a breakdown of the model based on binary
collisions, suggested for other reasons in Ref.~\cite{Gollub}.
It does however suggest a breakdown in any kinetic theory arguments based on
the Boltzmann equation which relies on the underlying assumption of 
molecular chaos and the statistical independence of collisions.  Similarly,
hydrodynamic descriptions which are based on a Chapman-Enskog expansion of
the Boltzmann distribution break down.  This is due to the breakdown in the 
statistical independence of collisions, not because ternary collisions are 
required to describe the system.
Thus, the collision time distribution can differentiate a granular fluid (exponential), a granular crystal (power law $\alpha=4$), and a granular {\it glass} (power-law $\alpha=2.8$).   

\begin{figure}
\includegraphics[width=14.25cm]{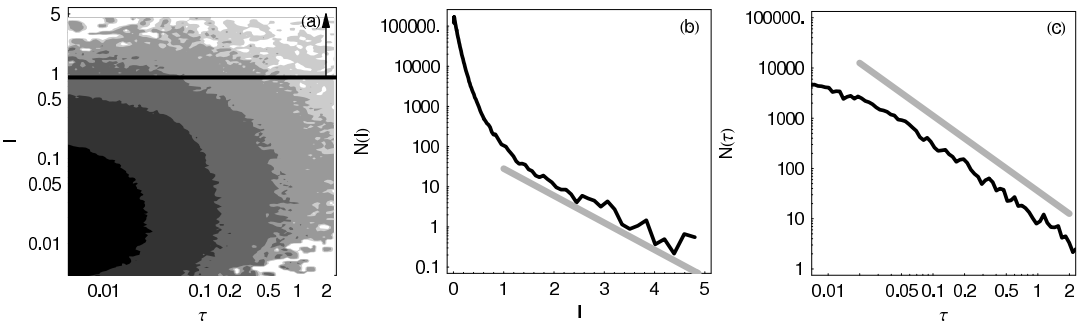}
\caption{
(a) Two-dimensional distribution of impulses and
collision times (log scales) for a $15\%$ ``spheres in 2D'' polydisperse
simulation.  The horizontal solid line in (a) indicates the cut-off in 
impulses, coinciding with the exponential tail
of the 1D impulse distribution shown in the semi-log plot (b), that results in the $1.5$ collision time power-law observed in (c).  Integrating the two-dimensional distribution in
(a) over all collision times gives the 1D histogram of impulses 
shown in (b).  Including only those impulses in 
the exponential tail of the impulse distribution (b), we obtain the $1.5$ collision time power law as shown by the thick gray line in (c).
}  
\label{fig5}
\end{figure}

Typically an exponential force distribution or impulse distribution is considered a signature of jamming.  It has been suggested however \cite{OHernandNagel,Ferguson,Silbert} that an upturn at the lower end of the impulse distribution may provide more information.
Impulse distributions from our spheres in 2D simulation, as shown in Fig.~\ref{fig5}(b), have a power law {\it upturn} at the low impulses and an exponential tail at the high impulses.  The power law at the lower impulses appears to be more a signature of jamming as even the granular fluid phase exhibits an exponential tail.

Our ``spheres in 2D'' simulation is a direct analog of the experiment described in \cite{Longhi}, the main difference being that the experiment discharged the grains from a hopper at the bottom of the chute whereas we use a sieving process at the bottom.  The experiment used a pressure transducer attached to the chute wall to detect grain-wall collisions, measure impulses, and times between collisions.    
However the experiment primarily sees an exponential distribution of impulses, although there is an upturn at small impulses for the slower flows.  Further,
for the distribution of grain-wall collision times they found a power-law with exponent $\alpha=1.5$.
Based on the distribution of impulses, we explain the main discrepancy not as a surface versus bulk effect but as a result of experimental response time and sensitivity of the detector as follows.  If we remove collisions with impulses in the power-law part of the distribution (i.e. the small impulse end that may be difficult to resolve experimentally) by putting in a cut off at low impulses we also get the $1.5$ collision time power law found in the experiment \cite{Longhi}, as shown in Fig.~\ref{fig5}.

In conclusion, power laws with exponents independent of dimension were found for the distribution of collision times.  Further, polydispersity and disorder are directly related to the power law exponent.   The collision time distribution is an exponential for a granular fluid, a power-law with exponent $\alpha=4$ for a granular crystal, and a power-law with exponent $\alpha=2.8$ for a granular {\it glass}.   
 It is interesting to note that the propagations of stress seems to show a similar ability to discern crystal and glass \cite{Behringer01}.  We successfully compared our collision time power law distribution with the experiment described in \cite{Longhi}.  By subtracting out the collisions with very small impulses that would be difficult to measure experimentally, we found that the stronger collisions that make up the exponential tail of the {\it impulse} distribution have an associated collision time distribution with power law exponent 1.5.  This highlights the importance of simulations that can give access to information not easily accessible in experiments.  Measurement of the full collision-time distribution in experiments would require accessibility to the measurement of very low impulse collisions.  In some systems this might be precluded by inelastic collapse and prolonged contacts.  However, groups have observed peaks in the low end of the force distribution for static systems\cite{OHernandNagel,Silbert} so it seems likely that a similar measurement of impulses in the analogous dynamic case could see the full spectrum of phenomenon we find in simulations. 

This work was supported by the Natural Science and Engineering Research Council of Canada, the Ontario Graduate Scholarship Program, and SharcNet.  We thank N. al Tarhuni for doing some preliminary simulations.


\begin{thebibliography}{99.}
\bibitem{polydispersity}
polydispersity = $x$\% $\Rightarrow$ standard deviation of particle radii is $0.x$ if mean is 1.
\end{thebibliography}
\end{document}